\documentclass[12pt]{article}

\usepackage{latexsym}

\begin{document}

\title{
Exact static solutions in four dimensional
Einstein-Maxwell-Dilaton gravity }

\author{
      S. Yazadjiev \thanks{E-mail: yazad@phys.uni-sofia.bg}\\
{\footnotesize  Department of Theoretical Physics,
                Faculty of Physics, Sofia University,}\\
{\footnotesize  5 James Bourchier Boulevard, Sofia~1164, Bulgaria }\\
}

\maketitle

\begin{abstract}
Classes of exact static solutions in four-dimensional
Einstein-Maxwell-Dilaton gravity are found. Besides of the well-known
solutions previously found in the literature, new solutions are
presented.It's shown that spherically symmetric solutions, except the
case of charged dilaton black hole, represent  globally naked
strong curvature singularities.
\end{abstract}


\sloppy
\renewcommand{\baselinestretch}{1.3} %
\newcommand{\sla}[1]{{\hspace{1pt}/\!\!\!\hspace{-.5pt}#1\,\,\,}\!\!}
\newcommand{\db}{\,\,{\bar {}\!\!d}\!\,\hspace{0.5pt}}
\newcommand{\partb}{\,\,{\bar {}\!\!\!\partial}\!\,\hspace{0.5pt}}
\newcommand{\dsla}{\partb}
\newcommand{\eql}{e _{q \leftarrow x}}
\newcommand{\eqr}{e _{q \rightarrow x}}
\newcommand{\ite}{\int^{t}_{t_1}}
\newcommand{\itz}{\int^{t_2}_{t_1}}
\newcommand{\itd}{\int^{t_2}_{t}}
\newcommand{\lfrac}[2]{{#1}/{#2}}
\newcommand{\dV}{d^4V\!\!ol}
\newcommand{\ben}{\begin{eqnarray}}
\newcommand{\een}{\end{eqnarray}}
\newcommand{\la}{\label}

\bigskip

The scalar field is believed to play an important role in physics.
Nearly all generalized theories of gravity as Brans-Dicke models and
Kaluza-Klein theory involve a scalar field. On the other hand the scalar
field arises from low energy limit of string theory.

In recent years a lot of attention has been turned to low dimensional
string theory inspirited models. It turns out to be useful to study such
models as a first step in the surch of the exact string theory.
On the one hand the studying of low dimensional models may throw light
on higher dimensional case and give some insight into the nonperturbative
structure of string theory.On the other hand it leads to better
understanding of some problems in pure general relativity - as example
black holes and the question of the scalar hair.
The charged black holes solutions supporting scalar hair in the four
dimensional Einstein-Maxwell-Dilaton  gravity  were found and studied in
\cite{GHS} (see also \cite{G}).

The purpose of the present article is to present classes of exact
static solutions in the four dimensional Einstein-Maxwell-Dilaton
gravity  containing new solutions besides of the known, previously
found in the literature.

We consider four dimensional Einstein-Maxwell-Dilaton gravity with
an action
\ben
S = {1\over 16\pi}\int d^4x \sqrt{-g}
\left(R - 2g^{ab}\partial_{a}\varphi\partial_{b}\varphi  -
e^{-2\varphi}F_{ab}F^{ab}\right)
\la{FA}
\een
$R$ being the Ricci scalar curvature with respect to the metric $g_{ab}$
and $F_{ab}=\partial_{a}A_{b}-\partial_{b}A_{a}$ being the Maxwell tensor.
The corresponding field equations are
\ben
\la{FE}
G_{ab} = 2\partial_{a}\varphi\partial_{b}\varphi -
g_{ab}\partial_{c}\varphi\partial^{c}\varphi  +
8\pi T_{ab}   \\ \nonumber
\Box\varphi = - {1\over 2}e^{-2\varphi}F_{ab}F^{ab} \\ \nonumber
\nabla_{b}\left(e^{-2\varphi}F^{ab}\right) = 0
\een
where $T_{ab}={1\over 4\pi}
e^{-2\phi}\left(F_{ac}F_{b}{}^{c} -
{1\over 4}g_{ab}F_{cd}F^{cd}\right).$
We look for static solutions of the field equations (\ref{FE}).This means
that we consider space-time admitting a Killing vector $\xi$ satisfying the
following conditions
\ben
g_{ab}\xi^{a}\xi^{b}<0 ,\,\,\,\,
2\nabla_{(a}\xi_{b)}={\cal{L}}_{\xi}g_{ab}=0
\,\,\,\, \xi_{[a}\nabla_{c}\xi_{b]}=0
\een
Besides the above conditions, the fields $\varphi$ and $F_{ab}$ are
required to satisfy
\ben
{\cal{L}}_{\xi}\varphi = 0 , \,\,\,\,\,\,\,\, {\cal{L}}_{\xi} F_{ab} = 0
\een

In the case of static space-time, as well-known, the metric can be written
in the form \cite{IW}
\ben
ds^2 = g_{ab}dx^{a}dx^{b}= -e^{2u}dt^2  + e^{-2u}h_{\mu\nu}dx^{\mu}dx^{\nu}
\een

In terms of the three-dimensional metric $h_{\mu\nu}$ the field equations
(\ref{FE}) become
\ben
\la{TFE}
{{}^{3}R}_{\mu\nu} = 2D_{\mu}uD_{\nu}u + 2D_{\mu}\varphi D_{\nu}\varphi  -
2e^{-2u -2\varphi}D_{\mu}\Phi D_{\nu}\Phi
\\ \nonumber
D_{\mu}D^{\mu}u = e^{-2u -2\phi}D_{\mu}\Phi D^{\mu}\Phi
\\ \nonumber
D_{\mu}D^{\mu}\varphi = e^{-2u -2\varphi}D_{\mu}\Phi D^{\mu}\Phi \\ \nonumber
D_{\mu}\left(e^{-2u -2\phi}D^{\mu}\Phi\right) = 0
\een
Here $D_{\mu}$ is Levi-Civita connection and ${{}^{3}R}_{\mu\nu}$ is the
Ricci tensor with respect to the three-metric $h_{\mu\nu}.$ The electric
potential $\Phi$ is introduced by
\ben
F_{ab} = -2e^{2u}\xi_{[a}\partial_{b]}\Phi
\een

The equations (\ref{TFE}) can be obtained from the following action
\ben
{\cal{A}} = \int d^3x \sqrt{h} \left({{}^{3}R} -
2h^{\mu\nu}\partial_{\mu}u\partial_{\nu}u  -
2h^{\mu\nu}\partial_{\mu}\varphi\partial_{\nu}\varphi +
2e^{-2u -2\varphi}h^{\mu\nu}\partial_{\mu}\Phi\partial_{\nu}\Phi\right)
\la{TA}
\een

Introducing the matrix \cite{EGK},\cite{GX}
\ben
P = \pmatrix{e^{2u} -2\Phi^2 e^{-2\varphi} & -\sqrt{2}\Phi e^{-2\varphi}\cr
-\sqrt{2}\Phi e^{-2\varphi} & - e^{-2\varphi}}
\la{P}
\een
the action (\ref{TA}) is written in the elegant form
\ben
{\cal{A}} = \int d^3x \sqrt{h} \left({{}^{3}R} - {1\over 2}
Sp\left(\partial_{\mu}P\partial_{\nu}P^{-1}\right) h^{\mu\nu}\right)
\een
Now we assume that all potentials $u,\varphi,\Phi$ depend on only one
potential $\lambda$ \cite{NK}.Then requiring that
\ben
{1\over 4}Sp\left({dP\over d\lambda}{dP^{-1}\over d\lambda}\right)=1
\la{NOR}
\een
the field equations (\ref{TFE}) are reduced to the following
\ben
\la{MFE}
{{}^{3}R}_{\mu\nu} = 2D_{\mu}\lambda D_{\nu}\lambda \\ \nonumber
D_{\mu}D^{\mu}\lambda =0  \\ \nonumber
{d\over d\lambda}\left(P^{-1}{dP\over d\lambda}\right)=0
\een
The first two equations of (\ref{MFE}) are actually the static vacuum
Einstein equations.The third equation can be formally solved.Its general
solution is
\ben
P = P_{0}e^{Q\lambda}
\een
where the matrices $P_{0}$ and  $Q$ must be determined.
From physical point of view, the most interesting case is when the
potential $\lambda$ vanish at the spatial infinity ($\lambda(\infty)=0$).
We are interested also in asymptotically flat solutions :
$u(\infty)=\varphi(\infty)=\Phi(\infty)=0.$
Then the matrix $P_{0}$ is determined by
\ben
P(\infty)=P_{0}
\een
Taking into account the explicit form (\ref{P}) of the matrix $P$ we obtain
\ben
P_{0}= \pmatrix{ 1 & 0\cr
0 & - 1} = \sigma_{3}
\een
Here $\sigma_{3}$ is the third Pauli matrix.

To determine the matrix $Q$ we note, as it's seen, that the matrix $P$ is
symmetric $P^{T}=P.$ Therefore Q must satisfies the condition
\ben
P_{0}Q = Q^{T}P_{0} \,\,\,\,\,\,\,
(or\,\,\,\,\,\, \sigma_{3}Q =Q^{T}\sigma_{3})
\la{CFQ}
\een
From (\ref{CFQ}) it immediately follows that
\ben
Q= \pmatrix{ \alpha & \beta  \cr -\beta   & \gamma}
\een
On the other hand the condition (\ref{NOR}) leads to the constraint
\ben
Sp(Q^2) = \alpha^2 + \gamma^2  - 2\beta^2 =  4
\een

The matrix $e^{Q\lambda}$ can be found by using standard technics.
In calculating $e^{Q\lambda}$  we must consider three cases :
$detQ < 2 $, $detQ > 2 $ and $detQ = 2 .$

First we consider the case $detQ < 2$. The convenient parametrization
in this case is
$\alpha + \gamma  = 2\sqrt{2}\sin(\omega),$
$\alpha - \gamma  =2\sqrt{2} \cosh(\Omega)\cos(\omega),$  and
$\beta = \sqrt{2}\sinh(\Omega)\cos(\omega).$

The following explicit form of $P$
is obtained
\ben
\la{1P}
P = e^{\sqrt{2}\lambda \sin(\omega)}
\pmatrix{\cosh(z) + \cosh(\Omega)\sinh(z)   &
\cosh(\Omega)\sinh(z)  \cr
\cosh(\Omega)\sinh(z)  &
\cosh(\Omega)\sinh(z) - \cosh(z)}
\een
where  $z = \sqrt{2}\lambda \cos(\omega).$

Now putting
\ben
\Gamma^2 = {\cosh(\Omega) -1\over \cosh(\Omega) + 1} \,\,\,\,\,\,
(\Gamma^2 < 1)
\een
it's not difficult to obtain
\ben
e^{2u} =\left(1 - \Gamma^2\right){e^{2\lambda \cos(\omega -{\pi\over 4})}
\over 1 - \Gamma^2 e^{2\sqrt{2}\lambda\cos(\omega)}} \nonumber
\een
\ben
e^{2\varphi} = \left(1 - \Gamma^2\right){e^{2\lambda
\cos(\omega +{\pi\over 4})}
\over 1 - \Gamma^2 e^{2\sqrt{2}\lambda\cos(\omega)}}
\la{FCS}
\een
\ben
\Phi = {\Gamma\over \sqrt{2}}
{1 - e^{2\sqrt{2}\lambda\cos(\omega)} \over
1 - \Gamma^2 e^{2\sqrt{2}\lambda\cos(\omega)}} \nonumber
\een

In the case when $det Q > 2$  we find
\ben
P = {e^{\sqrt{2}\lambda\cosh(\psi)} \over \cos(\vartheta)}
\pmatrix{ \cos\left(\sqrt{2} \lambda \sinh(\psi) - \vartheta \right) &
\sin\left(\sqrt{2} \lambda \sinh(\psi)\right)   \cr
\sin\left(\sqrt{2} \lambda \sinh(\psi)\right)   &
-\cos\left(\sqrt{2} \lambda \sinh(\psi) + \vartheta \right)}
\la{2P}
\een
where $\psi= arcosh({\alpha + \gamma \over 2\sqrt{2}})$ and
$\vartheta = \arcsin({\alpha - \gamma \over 2\beta}).$

From (\ref{2P}) one obtains
\ben
e^{2u} = e^{\sqrt{2}\lambda\cosh(\psi)}
{\cos(\vartheta) \over
\cos\left(\sqrt{2} \lambda \sinh(\psi) + \vartheta \right)} \nonumber
\een
\ben
e^{2\varphi} = e^{-\sqrt{2}\lambda\cosh(\psi)}
{\cos(\vartheta) \over
\cos\left(\sqrt{2} \lambda \sinh(\psi) + \vartheta \right)}
\la{SCS}
\een
\ben
\Phi = - {1 \over \sqrt{2}}
{\sin\left(\sqrt{2} \lambda \sinh(\psi)\right) \over
\cos\left(\sqrt{2} \lambda \sinh(\psi) + \vartheta \right)} \nonumber
\een

The third class solutions are obtained when $detQ = 2$ :
\ben
P = e^{\sqrt{2}\lambda}
\pmatrix{1 + {\alpha - \gamma \over \sqrt{2}}\lambda &
{\alpha - \gamma \over \sqrt{2}}\lambda  \cr
{\alpha - \gamma \over \sqrt{2}}\lambda  &
-1 + {\alpha - \gamma \over \sqrt{2}}\lambda }
\la{3P}
\een
Hence we find
\ben
e^{2u}= {e^{\sqrt{2}\lambda}\over
1 - {{\alpha - \gamma} \over \sqrt{2}}\lambda} \nonumber
\een
\ben
e^{2\varphi}= {e^{-\sqrt{2}\lambda}\over
1 - {{\alpha - \gamma} \over \sqrt{2}}\lambda}
\la{TCS}
\een
\ben
\Phi = -{{\alpha - \gamma}\over 2} {\lambda \over
1 - {{\alpha - \gamma} \over \sqrt{2}}\lambda} \nonumber
\een

Let's consider some explicit examples.If we take as a seed metric
the Schwarzschild solution
\ben
ds^2 = - e^{2\lambda}dt^2  +  e^{-2\lambda}dr^2  + r^2d\Omega^2
\een
where $\lambda = {1\over 2}\ln\left(1 - {2M\over r}\right)$ \, and choosing
the first class of the above described solutions, we find
\ben
e^{2u} = \left(1 -\Gamma^2\right)
{\left(1 - {2M\over r}\right)^{\cos(\omega - {\pi\over 4})}   \over
1 - \Gamma^2 \left(1 - {2M\over r}\right)^{\sqrt{2}\cos(\omega)}} \nonumber
\een
\ben
e^{2\varphi} = \left(1 -\Gamma^2\right)
{\left(1 - {2M\over r}\right)^{\cos(\omega + {\pi\over 4})}   \over
1 - \Gamma^2 \left(1 - {2M\over r}\right)^{\sqrt{2}\cos(\omega)}}
\la{SSS1}
\een
\ben
\Phi ={\Gamma \over \sqrt{2}}
{1 - \left(1 - {2M\over r}\right)^{\sqrt{2}\cos(\omega)}   \over
1 - \Gamma^2 \left(1 - {2M\over r}\right)^{\sqrt{2}\cos(\omega)}} \nonumber
\een
In the special case when $\omega = {\pi\over 4}$  we obtain  well-known
charged dilaton black hole solution \cite{GHS}, \cite{G}.

Another interesting  example is the choosing as a seed solution
the Curzon's metric \cite{Curzon}
\ben
ds^2 = - e^{2\lambda}dt^2  +
e^{-2\lambda}\left(e^{2K}(d\rho^2 + dz^2) + \rho^2d\Psi^2\right)
\een
with  $\lambda = - {M\over r}$, $2K = -{M^2\over r^2}\sin^2(\theta).$
In this case, choosing again the first class of solutions  (\ref{FCS})
we find
\ben
e^{2u} = \left(1 - \Gamma^2\right)
{e^{-2{M\over r} \cos(\omega - {\pi\over 4})}
\over 1 - \Gamma^2 e^{-2\sqrt{2}{M\over r}\cos(\omega)}} \nonumber
\een
\ben
e^{2\varphi} = \left(1 - \Gamma^2\right)
{e^{-2{M\over r} \cos(\omega + {\pi\over 4})}
\over 1 - \Gamma^2 e^{-2\sqrt{2}{M\over r}\cos(\omega)}}
\een
\ben
\Phi = {\Gamma\over \sqrt{2}}
{1 - e^{-2\sqrt{2}{M\over r}\cos(\omega)} \over
1 - \Gamma^2 e^{-2\sqrt{2}{M\over r}\cos(\omega)}} \nonumber
\een

Besides the above given explicit examples there are many possibilities
to generate new solutions using as seed solutions static solutions of the
vacuum Einstein equations.The  detailed study of all possible  solutions
exceeds the scope of this paper.We will investigate the spherically symmetric
solution (\ref{SSS1})  which is one of the most interesting.We will show that
solution (\ref{SSS1})  has a  globally  naked  strong
curvature singularity when  $\omega \ne {\pi \over 4}$.

The curvature singularity is usually found by showing the divergence of the
Kretschmann scalar at a finite affine parameter along a non-spacelike
geodesic.It's well-known that Ricci and Weyl scalars  are finite for
several type solutions having curvature singularities.In any case, the
divergence  of any of the above mentioned scalars shows the presence of
a curvature singularity.

The  Ricci and  Kretschmann  scalars for solution (\ref{SSS1}) are
correspondingly
\ben
R = 2(1 - \Gamma^2){M^2 \over r^4} {1\over
\left(1 -{2M \over r}\right)^{2 - \cos(\omega - {\pi\over 4})}} \\
\nonumber
{\left(cos(\omega + {\pi \over 4}) +
\Gamma^2\left(\sqrt{2}cos(\omega) - cos(\omega + {\pi \over 4})\right)
\left(1 -{2M \over r}\right)^{\sqrt{2}cos(\omega)}\right)^2 \over
\left(1 - \Gamma^2\left(1 -{2M \over r}\right)^{\sqrt{2}cos(\omega)}
\right)^3}
\een
\ben
R_{abcd}R^{abcd} =
4 e^{4u}\Biggl[
\left(u^{\prime\prime} + 2\left(u^{\prime}\right)^2 \right)^2  +
2\left(u^{\prime} - \lambda^{\prime} -
{1\over r}\right)^2 \left(u^{\prime}\right)^2  +  \\ \nonumber
2\left(u^{\prime\prime} - \lambda^{\prime\prime} +
\left(u^{\prime}  - \lambda^{\prime}\right)
\left(\lambda^{\prime} + {1\over r}\right) -
{\lambda^{\prime}\over r}\right)^2  +
\left({e^{-2\lambda} \over r^2}  -
\left(u^{\prime} - \lambda^{\prime} - {1\over r}\right)^2 \right)^2
\Biggr]
\een
where the prime denotes differentiation with respect to $r.$
These scalars diverge at $r = r_{min}$  demonstrating a curvature
singularity  there.Here $r_{min}$  is $2M$   when  $\cos(\omega)>0$
and $r_{min}$   is determined by
\ben
1 - \Gamma^2 \left(1 - {2M \over r}\right)^{\sqrt{2}\cos(\omega)} =0
\een
when $\cos(\omega)<0$.

A singularity is globally naked if there exists a future directed causal
curve  with one end on the singularity and the other end on the future
null infinity.To show that the curvature singularities in our case are
globally naked  we consider null geodesic in the
space-time described by  the  solution (\ref{SSS1}):
\ben
{dk^{a}\over d\tau}  + \Gamma_{bc}^{a} k^{b} k^{c} =0
\een
where $g_{ab}k^{a} k^{b}=0.$
The outgoing radial null geodesic  are
\ben
k^{0} =  Ce^{-2u} \\ \nonumber
k^{1} = C \\ \nonumber
k^{2} =k^{3} =0
\een
Here $C$ is an integration constant and $C > 0.$
It immediately  follows that
\ben
r = r_{\min}   + Ck
\een
and
\ben
dt = e^{-2u} dr .
\een
It can be verified that for finite $R$ we have
\ben
\lim _{\epsilon\to 0} \int_{r_{min} + \epsilon}^{R} e^{-2u}dr =
{1 \over 1 - \Gamma^2} \lim _{\epsilon\to 0} \int_{r_{min} + \epsilon}^{R}
{1 - \Gamma^{2}\left(1 -{2M\over r}\right)^{\sqrt{2}\cos(\omega)}
\over \left(1 -{2M\over r}\right)^{\cos(\omega - {\pi \over 4})}}dr
\nonumber
\een
\ben
\leq
\left\{\matrix{ R{\left({R\over 2M} - 1\right)^{1 -
\cos(\omega - {\pi\over 4})}
\over 1 - \cos(\omega - {\pi\over 4})} \,\, ,  &  \cos(\omega) \geq 0  \cr
R {1 \over
\Gamma^{2\left(1 - {\cos(\omega - {\pi\over 4})\over
\sqrt{2}\cos(\omega)}\right)}} \,\,,    &   \cos(\omega)<0}
\right.
\een
which are finite.Therefore the singularity $r=r_{min}$  is globally
naked.

A sufficient condition for a strong curvature singularity
is that at least along one null geodesic (with an affine  parameter $\tau$)
which terminates at singularity $\tau = 0$  the following is satisfied
\cite{TCE},\cite{CK},\cite{Joshi}
\ben
\lim_{\tau \to 0} {\tau}^2 R_{ab}k^a k^b \ne 0
\een

In our case we obtain
\ben
\lim_{\tau \to 0} {\tau}^2 R_{ab}k^a k^b  =
{1\over 2} \cos(\omega + {\pi\over 4})
\een
when $\cos(\omega) \geq 0$   and
\ben
\lim_{\tau \to 0} {\tau}^2 R_{ab}k^a k^b  =
{1\over 2}
\een
when $\cos(\omega) < 0.$
This shows that the globally naked singularity  is a strong
curvature singularity.

In a similar way it may be shown that second and third class spherically
symmetric solutions have globally naked strong curvature singularities.

In the present paper we have found classes of new exact static solutions
in the four-dimensional Einstein-Maxwell-Dilaton gravity.The well-known
charged dilaton black hole solution corresponds to the special case
$\omega = {\pi \over 4}.$

The spherically symmetric solutions have been studied in details.It has been
shown that when $\omega \ne {\pi \over 4}$ they have  globally
naked strong curvature singularities.In this way, it has been shown that
naked singularities arise in their own right in the four dimensional
Einstein-Maxwell-Dilaton gravity.This leads to the suspection that
cosmic censorship may be failed in dilaton gravity.Of course, this question
needs more complete and deeper investigation at least because  the
effective string action we have considered here is only a first
approximation in powers of $\alpha^\prime.$ In the vicinity of the
singularities the curvature is large and the higher order $\alpha^\prime$
corrections must be considered.
Nevertheless we hope that the results in this
paper may throw some light on the nonperturbative structure of string theory.

\bigskip
\noindent{\Large\bf Acknowledgments}
\bigskip

The author is grateful to P. Fiziev for his continuous encouragement
and valuable comments.

This work has been partially supported by
the Sofia University Foundation for Scientific Researches,
Contract~No.~245/99, and by the Bulgarian National Foundation
for Scientific Researches, Contract~F610/99.


\begin{thebibliography}{31}

\bibitem{GHS} D. Garfinkle, G. Horowitz, A. Strominger
                Phys. Rev. {\bf D43}, 3140 (1991); \\
                {\bf 45}, 3888(E) (1992)

\bibitem{G}  G.Gibbons Nucl.Phys. {\bf B207}, 337  (1982)

\bibitem{IW} W.Israel, G.Wilson  J.Math.Phys.(N.Y.) {\bf 13}, 865 (1972)

\bibitem{EGK} A.Eris, M. Gurses, A. Karasu J.Math.Phys.(N.Y.) {\bf 25},
                   1489 (1984)

\bibitem{GX}  M. Gurses, B. Xanthopoulos Phys. Rev. {\bf D26}, 1912 (1982)

\bibitem{NK} G.Neugebauer, D. Kramer Ann.Phys.(Leipzig) {\bf 24}, 62
                (1969)

\bibitem{Curzon} H. Curzon  Proc.London.Math.Soc. {\bf 23}, 477 (1924)

\bibitem{TCE}  F.Tipler, C. Clarke, G. Ellis  General Relativity and
                Gravitation,edited by Held (Plenum, NY, 1980)

\bibitem{CK} C. Clarke, A. Krolak  J. Geom.Phys. {\bf 2}, 127  (1986)

\bibitem{Joshi} P. Joshi, Global aspects in gravitation and cosmology,
                (Clarendon Press,Oxford, 1993)

\end{thebibliography}
\end{document}